\documentclass[
    ,final            
  ]
  {aipproc}

\layoutstyle{8x11double}
\usepackage{amsmath}

\begin{document}

\title{The Polyakov-Nambu-Jona-Lasinio model with six and eight quark interactions}


\author{J. Moreira}{
  address={Centro de F\'{\i}sica Computacional, Departamento de
         F\'{\i}sica da Universidade de Coimbra, 3004-516 Coimbra, 
         Portugal}
}

\author{B. Hiller}{
  address={Centro de F\'{\i}sica Computacional, Departamento de
         F\'{\i}sica da Universidade de Coimbra, 3004-516 Coimbra, 
         Portugal}
}

\author{A. A. Osipov}{
  address={Centro de F\'{\i}sica Computacional, Departamento de
         F\'{\i}sica da Universidade de Coimbra, 3004-516 Coimbra, 
         Portugal}
  ,altaddress={On leave from 
         Dzhelepov Laboratory of Nuclear Problems, 
         Joint Institute for Nuclear Research, 
         141980 Dubna, Moscow Region, Russia} 
}

\author{A. H. Blin}{
  address={Centro de F\'{\i}sica Computacional, Departamento de
         F\'{\i}sica da Universidade de Coimbra, 3004-516 Coimbra, 
         Portugal}
}

\begin{abstract}
The extension of the Nambu-Jona-Lasino Model with six and eight quark interactions to include the Polyakov loop is analysed. Several interesting features are determined to be an effect of the choice of the regularization procedure. The use of a Pauli-Villars covariant regularization enables a coherent and correct description of several quantities.
\end{abstract}

\maketitle


\section{The model}
The Nambu-Jona-Lasinio Model extended to include the 't Hooft determinant in the light quark sector (u,d and s) is regarded as a useful tool for the study of low energy hadron phenomenology as it shares with QCD the global symmetries, incorporates by construction a mechanism for dynamical chiral symmetry breaking and includes explicit breaking of the $U_A(1)$ symmetry. It has been shown that the inclusion of eight quark interactions can solve a fundamental problem of this model: the absence of a globally stable ground state \cite{Osipov:2005tq}. Several new features emerge from the addition of the 8q terms and are to a large extent dictated by the part of these interactions which violates the Okubo-Zweig-Iizuka (OZI) rule. At the same time they leave open the possibility of almost reproducing the same mesonic stectra \cite{Osipov:2006ns} that are obtained without them, thus preserving the validity of the model.
It was shown that the critical endpoint in the phase diagram of this model is pushed towards vanishing chemical potential and higher temperatures with increasing strength of the OZI-violating eight-quark interactions \cite{Hiller:2008nu}. The introduction of the eight quark interactions also affects the number of effective quark degrees of freedom \cite{Hiller:2008nu}. The correct asymptotic behavior of this thermodynamical quantity can be reproduced using a careful and consistent implementation of the Pauli-Villars regularization both in NJL and the PNJL models \cite{Hiller:2008nu}\cite{Moreira:2010bx}.

The extension to include the effect of the Polyakov loop can be done by introducing a background of static and diagonal (in the Polyakov gauge) temporal gluonic fields, $A_4=\imath A^0$, with $A^\mu=\delta^\mu_0 g A^0_{a}\frac{\lambda^a}{2}$ (where $\lambda^a$ are the GellMann matrices) coupled to the quark fields by the covariant derivative: $\partial^\mu\rightarrow D^\mu=\partial^\mu+\imath A^\mu$. The quantities $\phi$ and $\overline{\phi}$ (which in the framework of the model are treated as classical field variables) are then given by the trace of the Polyakov loop operator and its adjoint ($\mathcal{P}$ denotes path ordering, $N_c$ is the number of colors and $\beta$ is the inverse of the temperature): 
$\phi=\mathrm{Tr_c} L/N_c$, $\overline{\phi}=\mathrm{Tr_c} L^\dag/N_c$, $L=\mathcal{P} e^{\int^\beta_0 \mathrm{d}x_4 \imath A_4}.$

The transition between the confined-deconfined phases is driven by the temperature dependence of the additional pure gluonic term, the  Polyakov potential, $\mathcal{U}$, for which several forms have been proposed. In \cite{Ratti:2005jh} a polynomial form motivated by a Ginzburg-Landau \emph{ansatz} was used and we will refer to it as $\mathcal{U}^I$. In \cite{Roessner:2006xn} a form is used which we refer to as $\mathcal{U}^{II}$, including a logarithmic term inspired by the Haar measure of $SU(N_c)$ group integration. In \cite{Fukushima:2008pe} an exponential term derived in the strong coupling expansion of the lattice QCD action is included, $\mathcal{U}^{III}$, and in \cite{Bhattacharyya:2010wp} a form $\mathcal{U}^{IV}$, combines the polynomial and logarithmic forms (for details see these references and our discussion of these in \cite{Moreira:2010bx} where we also discuss the parametrization choice).
 
Integrating the gap equations selfconsistently with the stationary phase equations,
\begin{align}
\label{staeq}
\left\{
\begin{array}{l}
m_u-M_u=G h_u +\frac{\kappa}{16}h_d h_s +\frac{g_1}{4}h_u h^2_f+\frac{g_2}{2}h_u^3\\
m_d-M_d=G h_d +\frac{\kappa}{16}h_u h_s +\frac{g_1}{4}h_d h^2_f+\frac{g_2}{2}h_d^3\\
m_s-M_s=G h_s +\frac{\kappa}{16}h_u h_d +\frac{g_1}{4}h_s h^2_f+\frac{g_2}{2}h_s^3
\end{array}
\right. ,
\end{align}
(where $m_f$ and $M_f$ are the current and dynamical masses respectively), the thermodynamical potential is obtained
\cite{Hiller:2008nu} ($T$ is the temperature, $\mu$  the chemical potential, $G$, $\kappa$, $g_1$ and $g_2$ are the coupling strengths of the of the NJL, 't Hooft, OZI-violating and non-violating 8-quark interactions):
\begin{align}
\label{Omega}
&{\Omega\left(M_f,T,\mu,\phi,\overline{\phi}\right)}=\nonumber\\
=& \frac{1}{16}\left.\left(4Gh_f^2+\kappa h_uh_dh_s+\frac{3g_1}{2}\left(h_f^2\right)^2+3g_2h_f^4\right)\right|_0^{M_f} \nonumber \\
+&\frac{N_c}{8\pi^2}\!\sum_{f=u,d,s}\!\!\left( {J_{-1}(M_f^2,T,\mu,\phi,\overline{\phi} )} + C(T,\mu )\right)+ {\mathcal{U}\left(\phi,\overline{\phi},T\right)}.
\end{align}

The effect of the Polyakov loop was included straightforwardly by noting that its phase enters the action as an imaginary chemical potential thus resulting in the following generalizations of the usual expressions:
\begin{align}
\label{nfdefs}
\tilde{n}_{q(\overline{q})}\left(E_p,\mu,T,\phi,\overline{\phi}\right)&\equiv
\frac{1}{N_c}\sum_{i}^{N_c}n_{q(\overline{q})} (E_p,\mu+\imath \left(A_4\right)_{ii},T),
\nonumber\\
f^{+(-)}\left(E_p,\mu,T,\phi,\overline{\phi}\right)&\equiv Log \prod^{3}_{i=1}
\frac{e^{-\left(E_p\mp\mu\right)/T}}{n_{q(\overline{q})} (E_p,\mu+\imath \left(A_4\right)_{ii},T)}.
\end{align}
Here $E_p=|\overrightarrow{p}_E|^2+M^2$ and the (anti-)quark occupation numbers are give as usual by: 
$n_{q(\overline{q})}=(1+e^{(E_p\mp\mu)/T})^{-1}$

\section{Regularization}

We compare the results obtained using the usual 3-dimensional cutoff in the momentum integrals with a Pauli-Villars covariant regularization with two subtractions in the integrand. These are therefore specified by the action of the operators
\begin{align}
\hat{\rho}^{3D}&=\Theta\left(\Lambda-\left|\overrightarrow{p_E}\right|\right),\\
\hat{\rho}_{\Lambda\vec{p}_E}&=
 1-\left(1-\Lambda^2 \frac{\partial}{\partial \vec{p}_E^{\, 2}}\right)\exp\left(\Lambda^2\frac{\partial}{\partial\vec{p}_E^{\, 2}}\right),
\end{align}
in the momentum integrand function.

 The vacuum and medium contributions can be separated and depending on the choice of the regularization procedure we obtain for the vacuum contributions,
\begin{align}
&J^{vac (3D)}_{-1}(M^2)\nonumber\\
=&\Lambda\left(2\Lambda^3-\sqrt{M^2+\Lambda^2}\left(M^2+2\Lambda^2\right)\right)
+M^4 \mathrm{ArcSinh}\frac{\Lambda}{M}\nonumber\\
&J^{vac (PV)}_{-1}(M^2)\nonumber\\
=&\frac{M^4-\Lambda^4}{2}\ln(1+\frac{M^2}{\Lambda^2})
-\frac{M^2}{2}\left(\Lambda^2+M^2\ln\frac{M^2}{\Lambda^2}\right),
\end{align}
 and for the medium contributions we obtain:
\begin{align}
&J^{med (3D)}_{-1}(M^2,T,\mu,\phi,\overline{\phi})\nonumber\\
=&-\int^\Lambda_0\mathrm{d}|\overrightarrow{p_E}|~
8|\overrightarrow{p_E}|^2 T
(f^+_{M}+f^-_M-(f^+_0+f^-_0))\nonumber\\
&J^{med(PV)}_{-1}(M^2,T,\mu,\phi,\overline{\phi}),\nonumber\\
=&-\int^{\infty}_0\mathrm{d}|\overrightarrow{p}_E| \frac{8|\overrightarrow{p}_E|^4}{3}\hat{\rho}_{\Lambda\vec{p}_E}
\left(
\frac{n_{q~M}+n_{\overline{q}~M}}{E_p}-
\frac{n_{q~0}+n_{\overline{q}~0}}{|\overrightarrow{p}_E|}
\right).
\nonumber
\end{align}
Here we used the abbreviated notation with the subscripts $M$ and $0$ denoting the quantities defined in (\ref{nfdefs}) evaluated at those values for the mass (for the zero mass case we also set $\phi=\overline{\phi}=1$)  \cite{Hiller:2008nu}\cite{Moreira:2010bx}.

The constants of integration, $C(T,\mu)$, resulting from the integration of the gap equations over the mass are chosen as to counterbalance the part stemming from the zero-mass limit of integration and are therefore given by:
\begin{align}
 C^ {3D}(T,\mu)=&-\int^\Lambda_0\mathrm{d}|\overrightarrow{p_E}|8 |\overrightarrow{p_E}|^2 T 
 \left(f^+_0+f^-_0\right),\nonumber\\
C^{PV}(T,\mu) = &-\int^{\infty}_0\mathrm{d}|\overrightarrow{p}_E \frac{8|\overrightarrow{p}_E|^4}{3}
\left(\frac{n_{q~0}+n_{\overline{q}~0}}{|\overrightarrow{p}_E|}\right).
\end{align}
 Note that while this may not be the standard way used to derive the 3D result the final result is equivalent to the usual one
.
 
A common variation of this procedure has been to remove the cutoff of the medium parts as they are convergent. In the next section we compare results obtained using these four possibilities.

\section{Results}

Here we only present some selected results which illustrate the key points, for a more complete exposition and further details we refer to \cite{Moreira:2010bx}.

We tested the above mentioned four different regularization procedures with the four mentioned Polyakov potential forms, using parameter sets with and without eight quark interactions. The analysis of the normalized pressure difference at vanishing chemical potential which serves as a measure for the effective degrees of freedom,
$\nu(T)=(p(T)-p(0))/(\pi^2T^4/90)$, reveals that the failure to reach the Stefan-Boltzmann limit only happens in the case where we use the 3D cutoff everywhere (see Fig. 1a).

The removal of the cutoff in the medium part results in the dynamical mass dropping below the current mass going asymptotically 
to zero with the temperature increase (see Fig. 1b).

The removal of the cutoff leads to overshooting of the asymptotic solution dictated by the pure gluonic part in the case of potentials $\mathcal{U}^{I}$, $\mathcal{U}^{III}$ and $\mathcal{U}^{IV}$ (in $\mathcal{U}^{II}$ the logarithmic divergence prevents this from happening) as can be seen in Fig. 2a, 2b, 3a and 3b.  This overshooting does not occur if the cutoff is kept (see Fig. 2b for the finite chemical potential case).

The reason for these behaviours can be traced back to the asymptotic behaviour of the derivatives of $J_{-1}$ with respect to $M$, $\phi$ and $\overline{\phi}$ when $T\rightarrow\infty$: the derivative with respect to the mass diverges upon removal of the cutoff whereas with cutoff it goes to zero; the derivative with respect to $\phi$ ($\overline{\phi}$) diverges with $T^4$ upon the removal of the cutoff, the same order as the derivative of $\mathcal{U}$, originating a deviation from the solution dictated by the latter. The divergence is lower with cutoff and in this case the asymptotic solution is dictated by the Polyakov potential.

\begin{figure}[htp]
\centering
\begin{minipage}{\columnwidth}
$\begin{array}{cc}
\includegraphics[width= 0.49 \columnwidth]{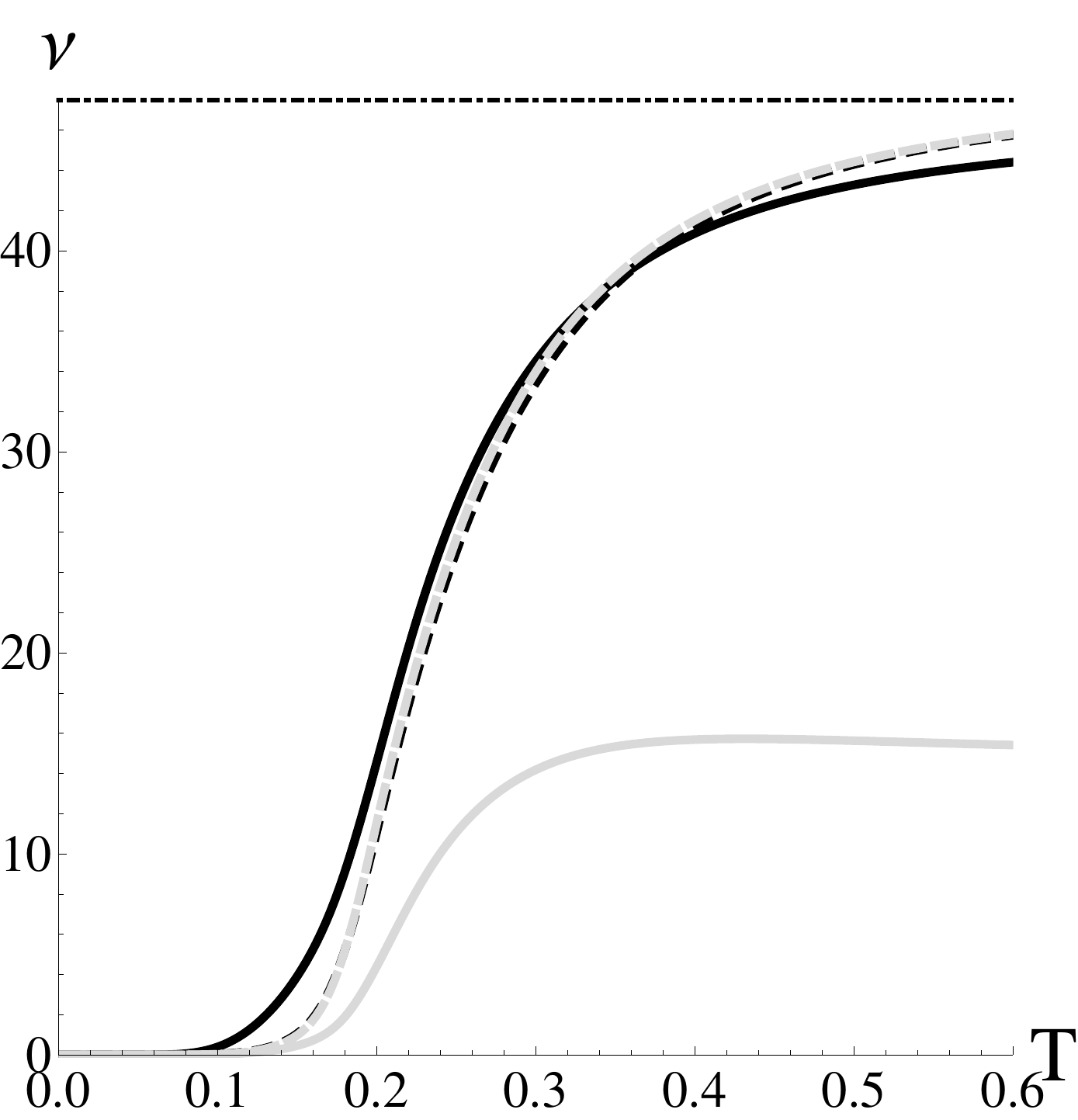}&
\includegraphics[width= 0.49 \columnwidth]{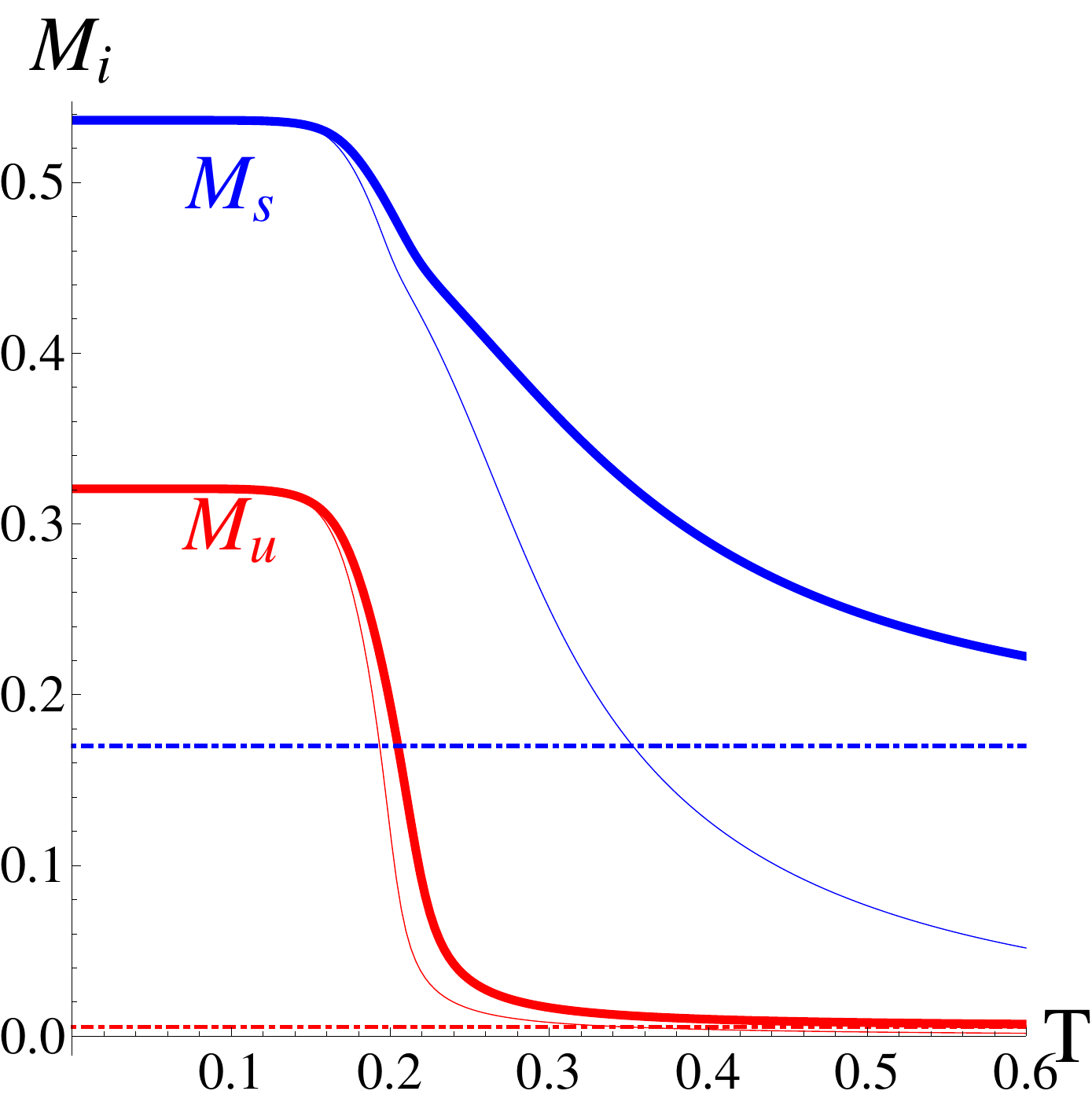}\\
 a & b
\end{array}$
\end{minipage}
\caption{Temperature dependence ($\left[T\right]=\mathrm{GeV}$) at $\mu=0$ in the PNJL model without eight quark interactions and with Polyakov potential $\mathcal{U}^{I}$ of: the number of effective degrees of freedom, 
$\nu(T)$, in 1a (black lines using PV and gray using 3D, dashing denotes the removal of the cutoff in the medium contributions), dynamical mass  of the quarks  ($\left[M_i\right]=\mathrm{GeV}$) using PV regularization 
(
thiner lines correspond to the removal of the cutoff in the medium). The results for the dynamical mass with 3D regularization are qualitatively similar.
}
\end{figure}

\begin{figure}[htp]
\centering
\begin{minipage}{\columnwidth}
$\begin{array}{cc}
\includegraphics[width= 0.49 \columnwidth]{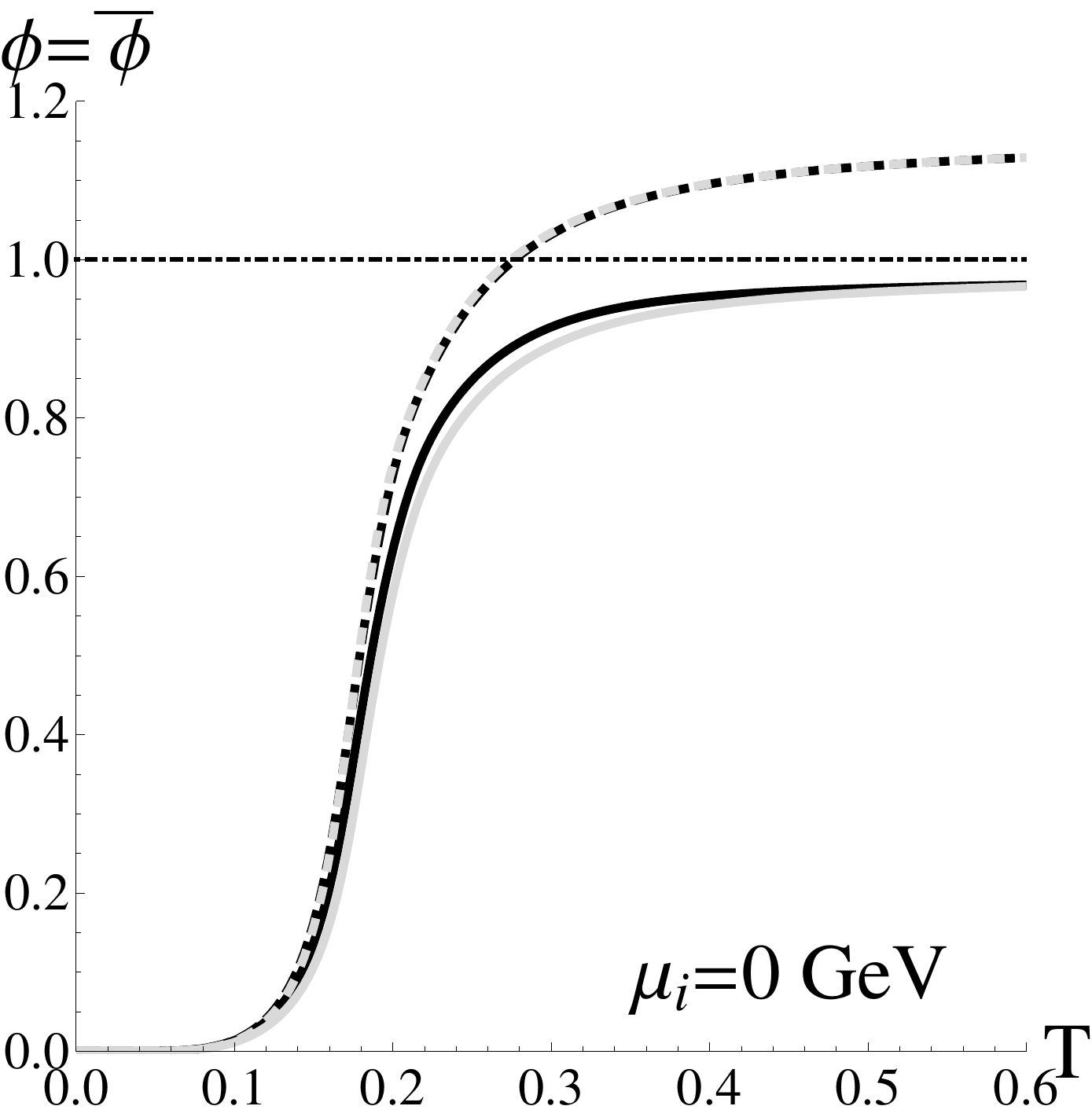}&
\includegraphics[width= 0.49 \columnwidth]{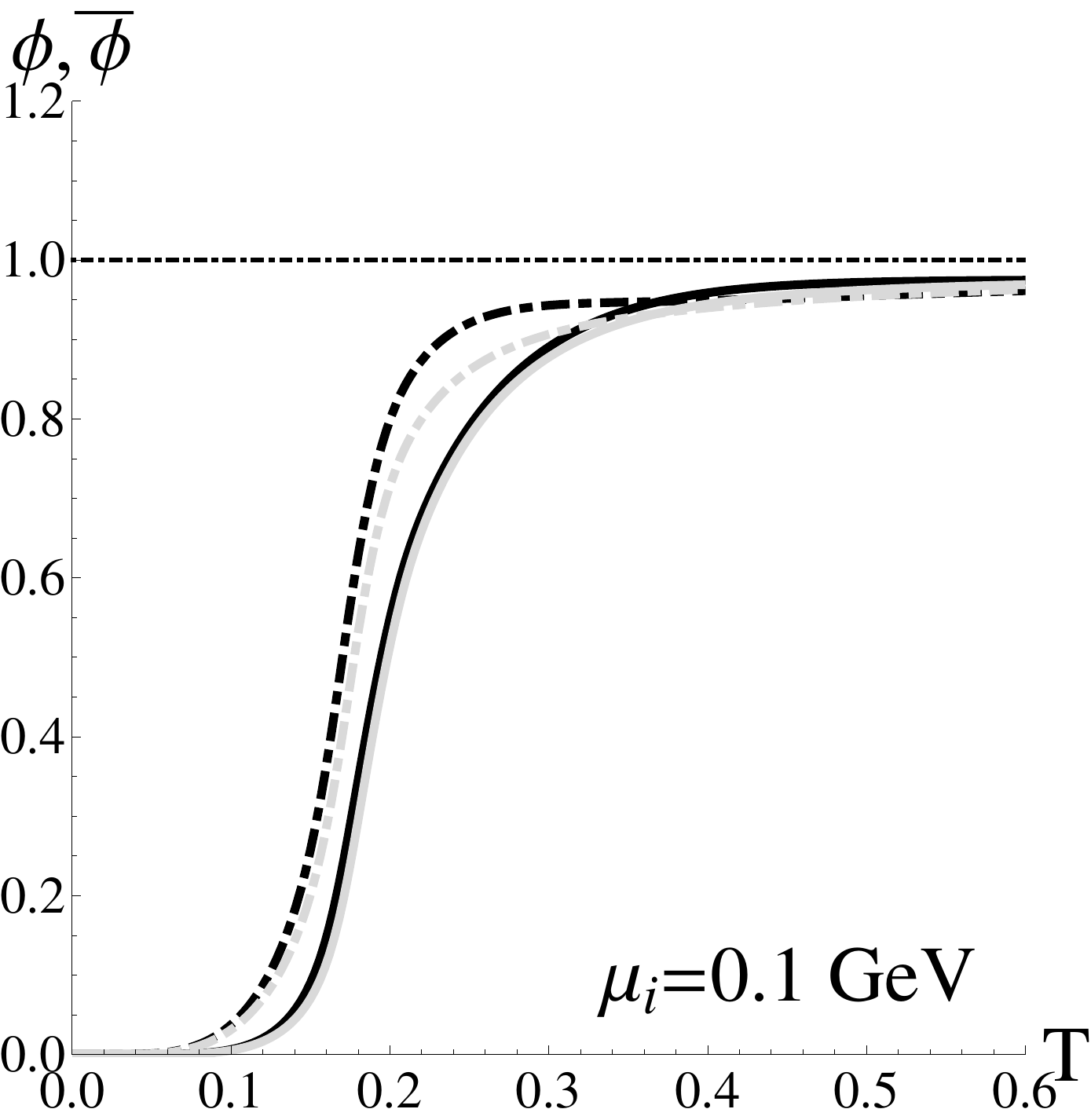}\\
a&b
\end{array}$
\end{minipage}
\caption{Temperature dependence of the Polyakov loop for the case withou 8q interactions and using $\mathcal{U}^{I}$: a) at $\mu=0$ ($\phi=\overline{\phi}$) the black lines refer to the use of PV and gray to 3D (dashed with only the vacuum regularized); b) $\phi$ in full lines and $\overline{\phi}$ in dashed lines at $\mu=100~\mathrm{MeV}$, black and gray lines refer to PV and 3D respectively (only the cases with cutoff everywhere are displayed).}
\end{figure}

\begin{figure}[htp]
\centering
\begin{minipage}{\columnwidth}
\includegraphics[width= 0.49 \columnwidth]{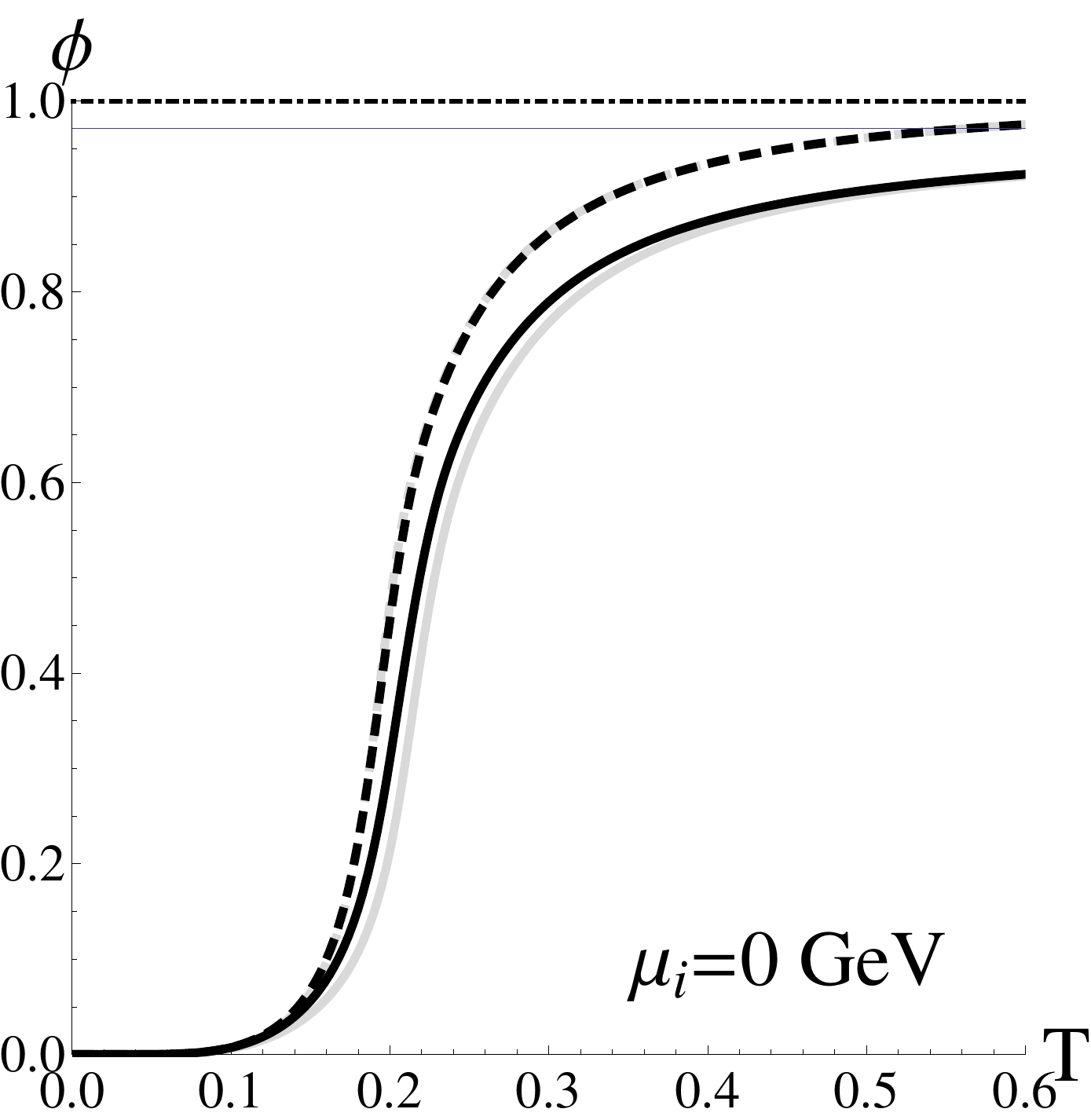}
\includegraphics[width= 0.49 \columnwidth]{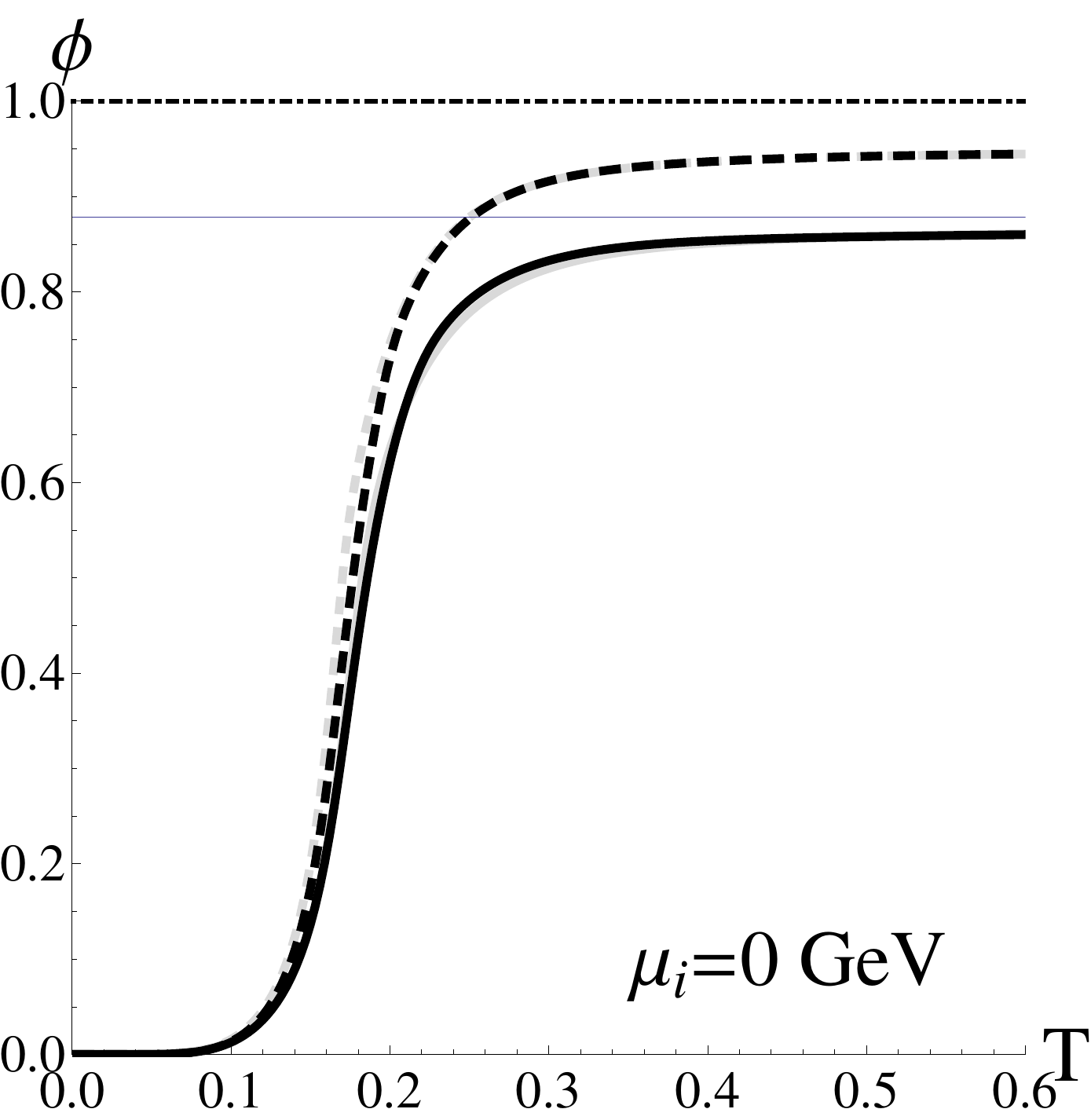}
\end{minipage}
\caption{Same as in Figure 2a but using $\mathcal{U}^{III}$ without 8q interactions in the left panel and $\mathcal{U}^{IV}$ with 8q interactions in the right panel (same color code for the regularizations as in 2a).}
\end{figure}



\section{Conclusions}
These qualitative features appear to be independent of the choice of parametrization (both of the quark interactions and the Polyakov potential) and are in fact a result of the regularization. The choice of Pauli-Villars regularization with the cutoff consistently kept over all contributions achieves the best results for the studied quantities.


\begin{theacknowledgments}
This  work  has  been  supported  in  part  by  grants  of Funda\c{c}\~{a}o para a Ci\^{e}ncia e Tecnologia,  FEDER,  OE,
SFRH/BPD/63070/2009, and Centro de F\'{i}sica Computacional, unit 405. We acknowledge  the support of the  European Community-Research  Infrastructure  Integrating  Activity  Study  of Strongly  Interacting  Matter  (acronym  HadronPhysics2, Grant   Agreement   No.   227431)   under   the   Seventh Framework Programme of the EU.
\end{theacknowledgments}



\bibliographystyle{aipproc}   

\bibliography{JMoreiraProceedings}

\IfFileExists{\jobname.bbl}{}
 {\typeout{}
  \typeout{******************************************}
  \typeout{** Please run "bibtex \jobname" to optain}
  \typeout{** the bibliography and then re-run LaTeX}
  \typeout{** twice to fix the references!}
  \typeout{******************************************}
  \typeout{}
 }

\end{document}